\documentstyle[12pt,epsfig]{article}

\begin{document}

\begin{titlepage}
\leftline{\_\hrulefill\kern-.5em\_}
\vskip -5 truemm
\leftline{\_\hrulefill\kern-.5em\_}
\centerline{\small{March 1994\hfill
Dipartimento di Fisica dell'Universit\`a di Pisa\hfill
IFUP-TH \ 16/94}}
\vskip -2.7 truemm
\leftline{\_\hrulefill\kern-.5em\_}
\vskip -5 truemm
\leftline{\_\hrulefill\kern-.5em\_}
\vskip1.5truecm
\centerline{Numerical simulation of}
\centerline{the Kardar-Parisi-Zhang equation}
\vskip.7truecm
\centerline{
${\rm Matteo\:Beccaria}^{\,1,\,2}$\ \
${\rm Giuseppe\:Curci}^{\,2,\,1}$}
\vskip1truecm
\centerline{\footnotesize{(1) Dipartimento di Fisica,
Universit\'a di Pisa}}
\centerline{\footnotesize{Piazza Torricelli 2, I-56100
Pisa, Italy}}
\vskip1truemm
\centerline{\footnotesize{(2) I.N.F.N., sez. di Pisa}}
\centerline{\footnotesize{Via Livornese 582/a, I-56010
S. Piero a
Grado (Pisa) Italy}}
\vskip1truecm
\begin{abstract}
\small{
We simulate the Kardar-Parisi-Zhang equation in $2+1$ dimensions.
It is
a non linear stochastic differential equation which describes
driven growing interfaces. The
Hopf-Cole transformation is used in order to obtain a stable
numerical scheme. The two
relevant critical exponents are precisely measured.
}
\end{abstract}
\vskip 1truecm
\hrule
\end{titlepage}

\vskip .5 truecm
\noindent
{\bf 1.\,Introduction}

Growing surfaces exhibit a non equilibrium critical dynamics with
scaling properties similar to those of equilibrium critical
phenomena.
Many models have been proposed to describe the universal
features of the
growth process. The first numerical investigations~\cite{Family}
showed that the critical behaviour of
the surface is described by two exponents $z$ and $\chi$. The
interface roughness $W$
(to be defined later) grows with time $t$ as
\begin{equation}
W\sim L^\chi\ f\left(t\ L^{-z}\right)
\end{equation}
where $L$ is the system size.
Among the possible systems possessing such a scaling law,
we mention for instance
lattice ballistic deposition
models, lattice stacking models or Eden clusters.
In~\cite{Kardar1}, Kardar, Parisi and
Zhang (KPZ) proposed a non linear stochastic
differential equation to describe growing interfaces.
Non linearity was related
to lateral growth of the interface.
The dynamic Renormalization Group
analysis~\cite{Kardar1,Medina,FNS}
determined the flow of the non-linearity parameter $\lambda$
related to the deposition speed.
The exponents $\chi$ and $z$ depend on the particular
fixed-point $\lambda^*$ which is
reached asymptotically. Dimensionality  of space is a priori
relevant.
If $\lambda^*\neq 0$, the hyperscaling relation $\chi+z=2$ holds
perturbatively.
In $d=1$ (growth on a line), the lowest order values for the
exponents
\begin{equation}
z = \frac{3}{2} \qquad \chi = \frac{1}{2}\qquad \beta =
\frac{\chi}{z} = \frac{1}{3}
\end{equation}
can be shown to be exact~\cite{Kardar1} and are supported
by several numerical simulations on the
various models listed above. In $d=2$, the
perturbative Renormalization Group analysis breaks
down, the flow is toward a strong coupling fixed-point
and numerical simulation becomes
very interesting.
Kardar, Parisi and Zhang conjectured~\cite{Kardar1} the
$\chi$, $z$ exponents to be superuniversal,
namely independent on $d$, but they did not give analytical
arguments. Successive works
on the subject can be roughly divided into three groups.
The first one deals with Eden cluster
growth~\cite{Devillard} and off lattice aggregation models
with possible
readjustement mechanisms~\cite{Jullien}.
The second group studies directed polymers~\cite{Kardar2}
and RSOS (Restricted Solid on Solid)
models~\cite{Forrest} which are
related to the KPZ equation. Finally, the third group
simulates directly the KPZ equation by
discretizing space and time and averaging over the
realizations of the
noise~\cite{Chackrabarti,Amar1,Grossmann,Moser}.

Apart from the direct simulations of the KPZ equation, the
information on the exponents
extracted from the other models is somehow indirect. Indeed,
directed polymers have been studied
only in the zero temperature limit whereas the other models
have not been
shown rigorously to be in the same universality class of the KPZ
equation.
On the other hand, the direct simulation of the stochastic
equation is hampered by great
crossover effects which are relevant on the time scale
actually explored in the simulations.
Numerical instabilities are also potentially harmful.

In Tab.(\ref{summary}), I have collected estimates for
$\chi$ and $\beta$ coming from
large scale simulations of Eden clusters~\cite{Devillard},
ballistic deposition models~\cite{Jullien},
directed polymers~\cite{Kardar2}, RSOS models~\cite{Forrest} and
direct simulation of the KPZ equation~\cite{Moser}(the $\chi$
exponent is taken from \cite{Amar1}).
I have also shown the $d=1$ exponents and the conjecture of
Kim and Kosterlitz~\cite{Kim}
\begin{equation}
\chi = \frac{2}{d+3}\qquad \beta=\chi/z = \frac{1}{d+2}
\end{equation}
based empirically on RSOS simulations in various dimensionalities.
\begin{table}
\begin{center}
\begin{tabular}{|c|cc|}
\hline
& $\chi$ & $\beta$ \\
\hline
Eden & 0.39(3) & 0.22(3) \\
Ballistic & $\sim 1/3$ & $\sim 1/4$ \\
Polymers & 0.53(7) & 0.33(2) \\
KPZ & $\sim 0.38$ & 0.240(5) \\
RSOS & 0.385(5) & 0.240(1) \\
\hline
KK & 2/5 & 1/4 \\
$d=1$ & 1/2 & 1/3 \\
\hline
\end{tabular}
\caption{Summary of the $d=2$ results.}
\label{summary}
\end{center}\end{table}
The situation in $d=2$ is therefore the following: (i) the
hyperscaling relation has strong numerical
support and (ii) the exponents seem to rule out the
superuniversality hypotesis and agree
reasonably well with the prediction of~\cite{Kim}.

Up to date, the most precise data on $\beta$, $\chi$ coming
from direct simulations of the KPZ
equation are those of~\cite{Moser} who however did not
check $\chi+z=2$. The authors
of~/cite{Moser} complain about numerical instabilities
arising at large non-linearity:~this unpleasant situation
forced them to utilize very small
integration steps not required in order to reduce the
systematic error due to finite integration step.
In this work we propose the simulation of the KPZ equation
after the
Hopf-Cole transformation which improves numerical stability
due to the elimination of the
non-linear term. We measure $\beta$ and $\chi$ with high
statistics and confirm the results
of~\cite{Moser}.

\vskip .5 truecm
\noindent
{\bf 2.\,The Renormalization Group analysis of the
Kardar-Parisi-Zhang equation}

The Kardar-Parisi-Zhang equation~\cite{Kardar1} in $d$
dimensions for the interface height
$h(x,\ t)$ is
\begin{equation}
\label{KPZ}
\frac{\partial h}{\partial t} = \nu\ \nabla^2 h +
\frac{\lambda}{2}\left(\nabla h\right)^2 + \eta(x,\ t)
\qquad x\in {\rm R}^d
\end{equation}
The noise $\eta$ is gaussian and the associated diffusion $D$
\begin{equation}
\langle\eta(x_1,\ t_1)\ \eta(x_2,\ t_2)\rangle = 2\ D\
\delta^{(d)}(x_1-x_2)\ \delta(t_1-t_2)
\end{equation}
is a constant.
The relaxation term in Eq.(\ref{KPZ}) provides the surface
tension responsible for
molecular readjustement. The non linear term is related to
lateral growing of the interface. The
KPZ equation is actually a truncated gradient
expansion:~therefore, full invariance under tilts is
lost. However, the infinitesimal symmetry
\begin{equation}
h\to h + \varepsilon\cdot x\qquad x \to x + \lambda\ t\
\varepsilon \qquad
\varepsilon\ {\rm infinitesimal}
\end{equation}
is unbroken. Indeed, if $h(x,t)$ is a solution, then also
\begin{equation}
h(x+\lambda\ t\ \varepsilon,\  t) + \varepsilon\cdot x
\end{equation}
is a solution up to $O(\varepsilon)$.
Let us introduce large scale variables
\begin{equation}
\tilde{x} = x/l\qquad \tilde{t} = t/l^z
\end{equation}
where $l$ is a length scale. Given a solution $h(x,\ t)$
corresponding to
parameters $(\nu,\  \lambda,\  D)$, the function
\begin{equation}
\tilde{h}(\tilde{x},\  \tilde{t}) = l^{-\chi}\ h(x,\ t) =
l^{-\chi}\ h (l\ \tilde{x},\  l^z\ \tilde{t})
\end{equation}
is a solution corresponding to new parameters $(\tilde{\nu},\
\tilde{\lambda},\  \tilde{D})$ where
\begin{equation}
\tilde{\nu} = l^{z-2}\ \nu \qquad \tilde{\lambda} = l^{\chi+z-2}\
\lambda \qquad
\tilde{D} = l^{z-d-2\chi}\ D
\end{equation}
Therefore, for $d>2$, we have the critical exponents
\begin{equation}
\label{naive}
z=2\qquad\chi = \frac{2-d}{2}
\end{equation}
and an asymptotically ideal surface with $\lambda=0$. To
explore $d\le 2$ we need a
one loop analysis. The evolution equations change into
\begin{eqnarray}
\label{pippo}
\nu^\prime(l)     &=& \left(z-2 + K_d
\bar{\lambda}^2\frac{2-d}{4d}\right)\ \nu \\
\lambda^\prime(l) &=& (\chi+z-2)\ \lambda \\
D^\prime(l)       &=& \left(z-d-2\chi+K_d\frac{1}{4}
\bar{\lambda}^2\right)\ D
\end{eqnarray}
where $\bar{\lambda} = \lambda^2 D/\nu^3$ and $K_d = S_d/(4\pi)^d$.
The second equation does not get perturbative corrections.
This is a non trivial consequence of the infinitesimal tilt
invariance.
It is a well known result of fluctuating
hydrodynamics~\cite{Medina}.
It follows that the hyperscaling relation
$\chi+z=2$ must hold at a non trivial critical point.
The running $\bar{\lambda}$ satisfies\footnote{
The evolution equation Eq.(\ref{pippo}) does not
change the naive results of
Eq.(\ref{naive}) since $\lambda=0$ is still stable.
However a transition at large $\lambda$
is possible in principle.
}
\begin{equation}
\bar{\lambda}^\prime = \frac{2-d}{2}\ \bar{\lambda} +
K_d\frac{2d-3}{4d}\ \bar{\lambda}^3
\end{equation}
The usual $\varepsilon$-expansion at the non-trivial
$O(\varepsilon)$ fixed point
gives at $d=1$ the exponents
\begin{equation}
d=1\qquad z = \frac{3}{2}\qquad \chi = \frac{1}{2}
\end{equation}
Furthermore, in $d=1$ a fluctuation-dissipation
theorem~\cite{Deker} can be invoked stating that at all orders of
perturbation theory the renormalizations of the evolution
equations of $\nu$ and $D$ are equal.
Therefore, the above exponents are exact.

However, at the critical dimension $d=2$, the flow is beyond
pertubation theory
with $\lambda$ growing as $l$ increases and going eventually
to an unknown
strong coupling fixed point.

\vskip .5 truecm
\noindent
{\bf 3.\,W and crossover}

The measure of $\beta$, $\chi$ is done by looking at the
interface roughness defined as
\begin{equation}
W = \langle h^2\rangle_c = \langle h^2\rangle -  \langle h\rangle^2
\end{equation}
where the average is over the lattice and $W(t, L)$ is
averaged over the noise.
At the unstable trivial fixed point $\lambda=0$, the
growth is marginal and
follows the law
\begin{equation}
W^2 \sim A\ \ln t\qquad
\end{equation}
Starting from $\lambda\neq 0$, we expect to observe the
asymptotic scaling
\begin{equation}
\label{scaling}
W\sim t^\beta
\end{equation}
in infinite volume. Finite size effects allow the observation of
Eq.(\ref{scaling}) only as an intermediate regime
and at $t\to\infty$ we have saturation with
$W\to W_{\rm sat}(L)$.
The determination of $\beta$ is rather difficult. If
$\lambda$ is too large, a small integration
step is needed to keep the systematic error small. If, on
the other hand, $\lambda$ is small,
crossover is observed with the scaling Eq.(\ref{scaling})
setting up slowly at small $\lambda$.
Moreover, saturation effects must always be kept under
control by comparing data obtained on larger and
larger lattices.

\vskip .5 truecm
\noindent
{\bf 4.\,Discretization and the Hopf-Cole transformation}

We rescale time and $h$ in order to reduce the number of
independent parameters
in the KPZ equation. Its canonical form is thus
\begin{equation}
\frac{\partial h}{\partial t} = \nabla^2 h +\sqrt{\lambda}
\left(\nabla h\right)^2 + \eta(x,\ t)
\end{equation}
with\footnote{
$\eta$ is a gaussian noise in the following discussion, but
in a lowest order
integration scheme, like the Euler one,
it can always be replaced by a uniform random number with
the same
first two momenta. The same trick is used in~\cite{Amar1}
where the invariance of the
results is explicitely checked.
}
\begin{equation}
\langle\eta(x_1,\ t_1)\ \eta(x_2,\ t_2)\rangle =
\delta^{(d)}(x_1-x_2)\ \delta(t_1-t_2)
\end{equation}
and where we have changed variable
\begin{equation}
\frac{2\lambda^2 D}{\nu^3} \to \lambda
\end{equation}
The equation depends thus on only one parameter which
determines the degree of non linearity.
The quadratic term $\left(\nabla h\right)^2$ is responsible
for numerical  instabilities since the typical surface becomes
more and more rough.
We propose to utilize the Hopf-Cole transformation
(see~\cite{Esipov} for a different application)
\begin{equation}
h = \frac{1}{\sqrt{\lambda}}\ \ln w \qquad w =
\exp\left(\sqrt{\lambda} h\right)
\end{equation}
At a formal level, this gives a diffusion equation
with multiplicative noise
\begin{equation}
\label{hopf}
\dot{w} = \nabla^2 w +\sqrt{\lambda}\ w\ \eta(x,\ t)
\end{equation}
In principle, there are ambiguities in the
discretization of~Eq.(\ref{hopf}) because the noise is
not differentiable.
To be general, let us consider the following stochastic
equation for $h(x,\ t):R^d\times R\to R$
\begin{equation}
\label{basic}
\dot{h} = f(h) + \xi\qquad \langle \xi(t_1)\xi(t_2)\rangle =
\delta(t_1-t_2)
\end{equation}
Its precise interpretation is the discrete process
\begin{eqnarray}
h_{n+1} &=& h_n + \varepsilon f(h_n) + \sqrt{\varepsilon}\ \xi_n \\
\xi_n &:& \mbox{normal random variables}
\end{eqnarray}
and averages over the stochastic realizations must be
extrapolated to the
$\varepsilon\to 0$ limit.
The result of a general transformation of $h$ is well
understood in terms of the
corresponding Fokker-Planck equations. If $P_h(h,t)$
is the density of processes relative to
Eq.(\ref{basic}), we have
\begin{equation}
\label{FP}
\frac{\partial P_h}{\partial t} = \frac{1}{2}
\frac{\partial^2 P_h}{\partial h^2} -
\frac{\partial}{\partial h} \left(f(h) P_h\right)
\end{equation}
We now consider the change of variables $h = \ln w$
(the following reasoning is however
completely general, provided that the change of variables
is well defined).
$P_h$ is a density and transforms accordingly as
\begin{equation}
P_h = \left|\frac{\partial w}{\partial h}\right| P_w = w P_w
\end{equation}
Substituting into Eq.(\ref{FP}) we obtain
\begin{equation}
\label{FPP}
\frac{\partial P_w}{\partial t} = \frac{1}{2}
\frac{\partial^2 }{\partial h^2}(w^2 P_w) -
\frac{\partial}{\partial h} \left[w(f(\ln w) + 1/2) P_w\right]
\end{equation}
A simple discrete process which in the $\varepsilon\to 0$
limit gives rise to Eq.(\ref{FPP}) is
\begin{equation}
w_{n+1} = w_n + \varepsilon\ w_n\ \left[f(\ln w_n) +
\frac{1}{2} \right] +
\sqrt{\varepsilon}\ w_n\ \xi_n
\end{equation}
The same result is obtained by applying directly the
Hopf-Cole transformation to the Euler
discretization of the original equation. Here, the additional
term in the force
is not $1/2$, but $1/2 \xi_n^2$; of course, there is no difference.

We remark that we have eliminated all the potential
ambiguities related to the choice of
Ito or Stratanovitch calculus since we have constructed
explicitely our stochastic processes,
by their unambiguous discrete realizations.

\vskip .5 truecm
\noindent
{\bf 5.\,The simulation}

The numerical simulation has been done on the {\sf APE}
supercomputer~\cite{ape:machine}.
The model operating at Pisa is the so called ``{\sf tube}'' machine,
a 128 processor parallel computer with a peak performance
of 6 GigaFlops.
We have simulated the $d=2$ growth process on a square
lattice with sizes up to $L=512$.
Every integration of the KPZ equation was averaged over
128 realizations of the noise.
The value of the integration step was chosen in order to
have irrelevant differences with
smaller values. In Fig.(\ref{fig:beta}), we show (from
top to bottom)
the behaviour of $W$ at $\lambda = 25, 10, 7.5, 5, 2$; we have used
$\varepsilon = 5\cdot 10^{-4}, 10^{-3}, 2.5\cdot 10^{-3}$
for the first three and
$\varepsilon = 5\cdot 10^{-3}$ for the others. At
$\lambda = 25$ we have taken data on the
$L=512$ lattice, the other values are at $L=256$.
The crossover effect inducing effective exponents is evident
in the figure. In order to avoid finite
size effects, we determined $\beta$ over a range where
changing $L$ from $256$ to $512$ was irrelevant.
We obtained $\beta = 0.240(1)$. The exponent $\chi$ has
been obtained by studying the saturation
width as a function of $L$ by mean of long runs. The
points in Fig.(\ref{fig:chi})
give $\chi = 0.404(1)$. Therefore, the hyperscaling relation
is well satisfied. Our
results are compatible with~\cite{Moser} and with the
empirical conjecture of~\cite{Kim}.
Of course, we cannot be sure to have eliminated totally
the crossover effects. The common lore is
that, due to saturation effects, numerical estimates for
$\beta$ must be considered lower bounds.
It is clear that, even if the agreement with RSOS models
is encouraging, an analytical upper
bound would be desirable.

\vskip .5 truecm
\noindent
{\bf 7.\,Conclusions}

The main goal of this paper has been an high statistics
simulation of the KPZ
equation put in an alternative form after the Hopf-Cole
transformation.
This change of variables in the stochastic equation does
not present any problem from the
point of view of the simulations and is numerically stable.
Our measures of $\beta$ and $\chi$ satisfy the hyperscaling
relation and are compatible with the conjecture of~\cite{Kim}.

\vskip .5 truecm
\noindent
{\bf Acknowledgements}

We thank Prof. Raffaele Tripiccione for a continuous help in
developing
the codes for the {\sf APE}
machine.

\psdraft

\begin{figure}[htbp]
\begin{center}
\mbox{\psfig{file=beta.ps,width=3.0truein,angle=-90}}
\end{center}
\caption{$W$ versus $t$ at $\lambda = 25, 10, 7.5, 5, 2$}
\label{fig:beta}
\end{figure}

\begin{figure}[htbp]\begin{center}
\mbox{\psfig{file=chi.ps,width=3.0truein,angle=-90}}
\end{center}
\caption{Determination of $\chi$}
\label{fig:chi}
\end{figure}

\end{document}